# Data Acquisition Systems and Computing Infrastructure of the IBR-2 Complex: Present and Near Future

A.S.Kirilov, F.V.Levchanovski, V.I.Prikhodko

Joint Institute for Nuclear Research, 141980 Dubna, Moscow reg., Russia

### 1. Introduction

In the period 2003-2009 in the Frank Laboratory of Neutron Physics, JINR, the total modernization of the IBR-2 pulsed reactor, which is mainly used for investigations in condensed matter physics using thermal neutron scattering will be carried out (it is planned to replace the movable and stationary reflectors and reactor jacket, to conduct fuel reloading, to update the automated system for measurement and control of the reactor parameters, etc). The main aims of the IBR-2 modernization are to improve the reactor parameters and to increase the nuclear safety and reliability of the reactor. In the mentioned period the development of new instruments and modernization of the existing neutron spectrometers are planned as well. In this connection in FLNP the 7-years program of the development of DAQ system and computing infrastructure was worked out.

The main problems in building new DAQ systems for neutron spectrometers are reliability, cost, time of development and commissioning and rapid system adaptation. Adaptability is of key importance since it gives the flexibility to meet unknown future needs.

Since in multi-layer architecture of DAQ systems the only part, which can be easily modified is software, it is necessary to begin the digital signal processing as early as possible. It was precisely this approach that was used when designing unified VME DAQ systems now in service at the IBR-2 spectrometers.

### 2. VME data acquisition and control systems

During the last years practically on all the IBR-2 spectrometers the measuring and control systems were changed over to a new generation of electronics, computers and software on the basis of VME systems and workstations integrated into the FLNP local area network [1].

**Data acquisition systems**

All digital electronics is placed in VME crates, equipped with standard purchased modules (processor E17 with a memory module, disk subsystems, input/output registers, ADC/DAC module, network interface, RS232 serial ports) as well as with modules developed in FLNP, which are specific for neutron time-of-flight experiments. The digital systems for registering and accumulating data from different detectors of the instruments represent a unified set of identical (from the viewpoint of hardware) modules limited in type but functionally complete. The systems realize distinction in parameters, functional capabilities, encoding, correction and preliminary data processing procedures, that are specific to each spectrometer on the level of microprograms, electronic tables, switches, etc.

For any IBR-2 spectrometer the data acquisition and accumulation system is equipped with the following basic modules (Fig.1): interface modules of position-sensitive detectors, module for encoding the point detector number, processor module, histogram memory and RTOF-analyzer.

The main part of DAQ system is located in the processor module (based on the high-performance DSP 'C40) which is responsible for ping-pong data buffering for each cycle of the reactor, data format transformations, calculating neutron position codes in linear and two-coordinate position-sensitive detectors, computing neutron TOF correction (time focusing), calculating addresses for the histogram memory, as well as for controlling the parameters of the reactor and spectrometer, different fault conditions and executing commands of the VME system.

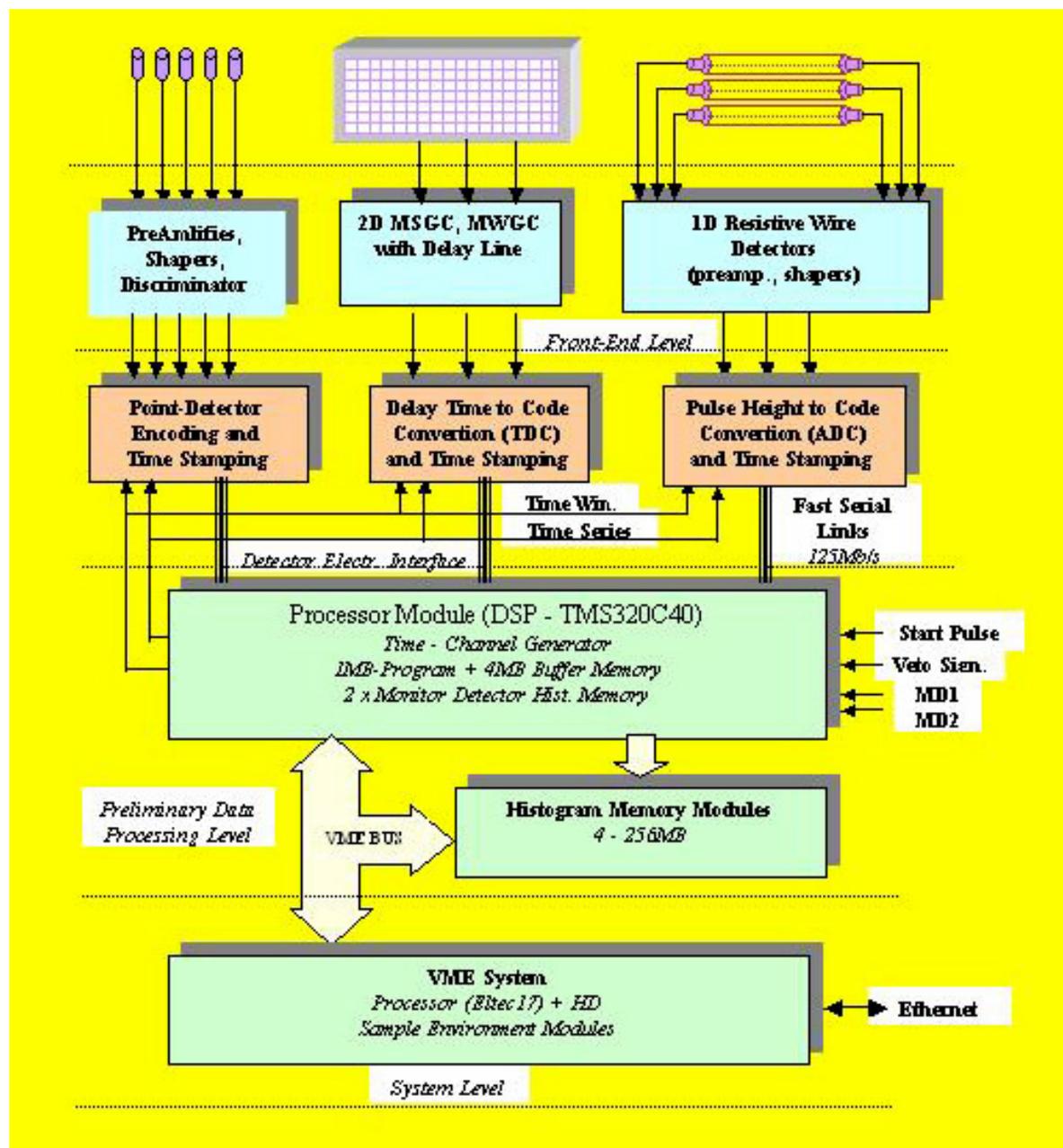

*Fig. 1. Data acquisition architecture.*

**Instrument and sample environment control systems**

In the specified spectrometers a rich variety of equipment for positioning samples and detectors (goniometers, rotary platforms, step motors), and for changing conditions on a sample (temperature, pressure, magnetic field, etc.) is used. Control systems of this equipment are also unified and consist of the following subsystems:

− Control systems for executive mechanisms (step motors, etc.);
− Regulation systems based on the Euroterm, LTC or DRC regulators (temperature);
− Systems for acquisition of analog parameters from sensors of various types;
− Systems for controlling logical parameters.

**Software**

On all the spectrometers the standard software SONIX [2] developed in FLNP is used. The SONIX programs are run by the VME processor in the real time operating system OS-9. The SONIX has a modular structure and includes the following functional blocks:

| Name | Module function |
|---|---|
| Join | Script interpreter |
| Tofa | DAQ control |
| Unipa | Control over a neutron beam and sample environment parameters |
| Motor | Control over goniometers and other step-motor driven devices |
| Temp | Control over furnaces/refrigerators |
| Graph | Visualization protocol (temperature, etc.) |
| Vsp | Visualization of one-dimensional spectra (on-line and off-line) |
| MAX | On-line visualization programs (1D, 2D) |

The tasks specified in the table are configured according to the requirements of concrete spectrometers. The OpenG2 programs based on the PV-WAVE commercial package are adapted for visualization of spectra from one and two-coordinate detectors. These programs provide preliminary processing, analysis and visualization of experimental data.

As noted above, the electronics and software of data acquisition and control systems are chosen from a set of unified electronic and program modules, depending on the equipment composition of spectrometers, type and number of detectors, and the required parameters (number and width of time channels, delays relative to the reactor startup, temperature range, etc.) are programmed or set at the level of blocks (DSP) or SONIX program modules.

During the experiment, data (spectra) are accumulated in histogram memory, written on a disk of a VME-computer and are transferred to the SUN-cluster for preliminary processing and visualization. The control over the experiment can be exercised from any computer of the FLNP local network.

**Development of DAQ and control systems**

Using the same principle, DAQ systems for future spectrometers and experiments at the modernized IBR-2 reactor, will be designed.

The main directions of these works:

- Integration of PCs into the structure of VME DAQ systems.
- Integration of PCs into separate measurement and control subsystems and creation of PC mini-farms for controlling spectrometers and data processing.
- Development of new, specific for neutron experiments, electronic modules with improved characteristics with due regards to natural progress in designing and manufacturing of electronic components. This will make it possible:
— to increase the counting rate of DAQ systems up to $10^7$-$10^8$ events/s;
— to extend the histogram memory volume up to tens of G bytes ;
— to improve the accuracy of measurements and to increase the number of registered and controlled parameters;
— to use more sophisticated algorithms for correcting and processing data, etc.
- Galvanic isolation of detector electronics.
- Reduction of the number of our own developments and orientation to a wide application of industrial standards and systems, as well as commercial software products.

- Application of new technologies for designing DAQ systems, primarily mezzanine and network technologies.

At present, works are carried out in all specified directions, two of them will be considered in more detail.

**Use of VME-PCI adapters**

Though even now VME systems operate successfully, moral aging of computers (processor module E17 based on Motorola 68040 processor) and slow development of the OS-9 operating system call for modernization of the existing systems. This is especially true for the spectrometers with position-sensitive detectors, where data flow is an order of magnitude higher. Unfortunately, the replacement of VME processor blocks by modern analogues also demands the upgrading and software development tools, which is very expensive in the aggregate.

The approach that has been accepted in a number of Institutes makes it possible to achieve new quality and find a cheaper direction for the future modernization. The main idea of this solution consists in replacing the VME processor block by an external computer (PC or others) connected to the VME crate using VME-PCI adapter. Upon this manipulation, the VME crate from an independent computer is turned into a complex controller, control over which is completely transferred to the external computer. This approach allows one to retain already created hardware blocks in the VME standard, though it requires the renewal of software. If a PC with Windows operating system is chosen as an external computer, the difficulties connected with the replacement of software are significantly minimized, and the user has the possibility to work in habitual, friendly environment, rich in various useful software products.

The change of the software platform will also permit us to develop significantly the Sonix software complex, which was hampered by a limited number of means available in the Laboratory for developing software in the OS-9 environment. The use of Windows instead of OS-9 will make it possible:

- to remove a barrier between data processing programs on the basis of PV-WAVE, IDL packages and control systems of spectrometers;
- to give rise to wide introduction of modern data formats (Nexus, XML, etc.);
- to apply the advanced programming technologies;
- to facilitate the inclusion of programs for primary data processing into the structure of spectrometer software, which were developed by physicists for PC or borrowed in other neutron centers.

For the centralized storage of experimental data from the IBR-2 spectrometers it is proposed to organize a special server based on a powerful computer on the Intel platform with disk RAID-array. Files with all measured data will be automatically transferred to this server from control computers of the spectrometers. The server is intended for long-term storage of spectra and to provide shared access to them, including from outside the Laboratory. The server also can be used to publish periodically updated information on current state of measurements at the spectrometers in WWW.

For organization of remote control over measurements and remote access to the spectrometers via Internet, the scheme used as the basis for the BARNS system [3] operating in ILL, and for the IBR-2 monitoring systems [4] available in FLNP, is proposed. On the dedicated computer a special-purpose web-server is installed, which provides periodically updated information on the state of measurements in Internet. By users' request this server contacts the file server and generates a web page with the spectrum of interest and/or state of measurement on a concrete spectrometer.

The first experience in using the VME-PCI adapter (Model 617 by the SBS firm) [5] was gained on the NERA-PR spectrometer. This system of alternative control that was in service for three last cycles of the IBR-2 in 2002 demonstrated high stability. In the future similar works will be carried out for other spectrometers as well.

**Integration of PC into detector system**

In cooperation with HMI, Berlin, a new version of the unified TDC/DSP DAQ block for acquisition and accumulation of data from position-sensitive detectors (multiwire proportional chamber with delay line data readout), was developed and manufactured (Fig. 2). It comprises 8-channel TDC of F1 type (Acam), FIFOs of different types, CPLDs, 256 Mbyte histogram memory and DSP TMS 320C6711. In the block, the determination of X/Y coordinates of the event (by signals from both ends of the delay lines) and neutron time of flight from the reactor start to the moment of detection, is executed; for the methodical purposes the amplitude of signals is measured as well. Two main operating modes are provided: histogram (on-line sorting of data and building of spectra) and "list" (accumulation of raw data with subsequent off-line processing). It is also possible to simultaneously accumulate histograms (for experiment control) and to write raw data. The TDC/DSP block has a PCI-interface and is installed directly in the case of PC. At present, the adjustment of electronics and debugging of microprograms (DSP) of the block are under way.

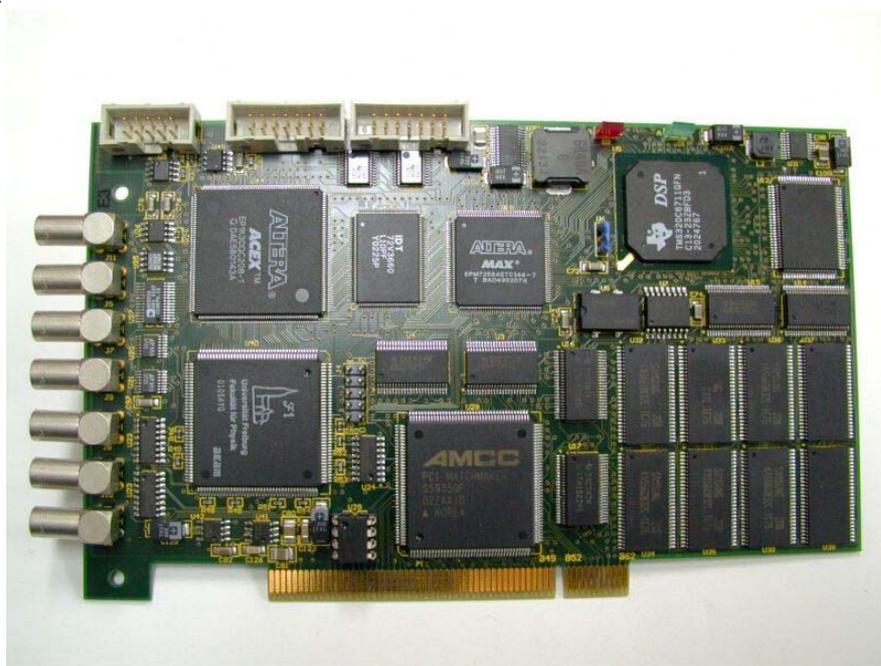

*Fig. 2*. TDC/DSP block with PCI interface.

For this block the architecture was developed and the debugging of the prototype of the program driver was carried out. The driver provides interaction between the program modules of the low (DSP) and following (PC) levels for several variants of basic software packages: C++, PV-WAVE, ROOT.

The software should be realized as a complex of independent program modules that includes a set of programs for the DSP processor, which can be loaded directly into the program memory of the block, program driver of the TDS/DSP block, program modules of intermediate layer and program applications providing a full set of functions to control operation of the block, to read and visualize the accumulated data.

Thus, the detector system is created, which will include the detector, detector electronics, registration and data accumulation subsystem, PC and software. In essence it will be complete autonomous elements of spectrometers that are built in any experiment control system.

### 3. FLNP information and computing infrastructure

Now the FLNP LAN is characterized by the following parameters:

- number of nodes > 500
- data transfer rate – up to 100 Mbit/s (via optical and twisted pair cables)
- number of workstations and servers – 20 (including SUN Sparc, Ultra SUN Sparc, Enterprise 3000)
- services – WWW, e-mail, databases, general purpose software, etc.

A historically established star-like communication structure (8 buildings within a radius of 1 km) and the absence of intellectual control devices do not make it possible to distribute flows between users in an optimal way. In addition, the transition of the experiments conducted with the FLNP facilities to a distributed model of data storage and processing also increases load on LAN. The possibility of experiment control imposes much stricter requirements to such network parameters as package transportation time and percent of losses.

In 2001, a long-term project of modernization of infrastructure of LAN basic segments and key elements was developed. At the first stage of the project (2001-2003) in LAN the router of information flows CISCO 8510 (with interfaces for optical communication lines between buildings and twisted pairs to connect SUN-cluster servers) was put into operation. In all buildings high-speed switches Catalist 29XX (CISCO) for communication with the network central segment were installed (Fig. 3).

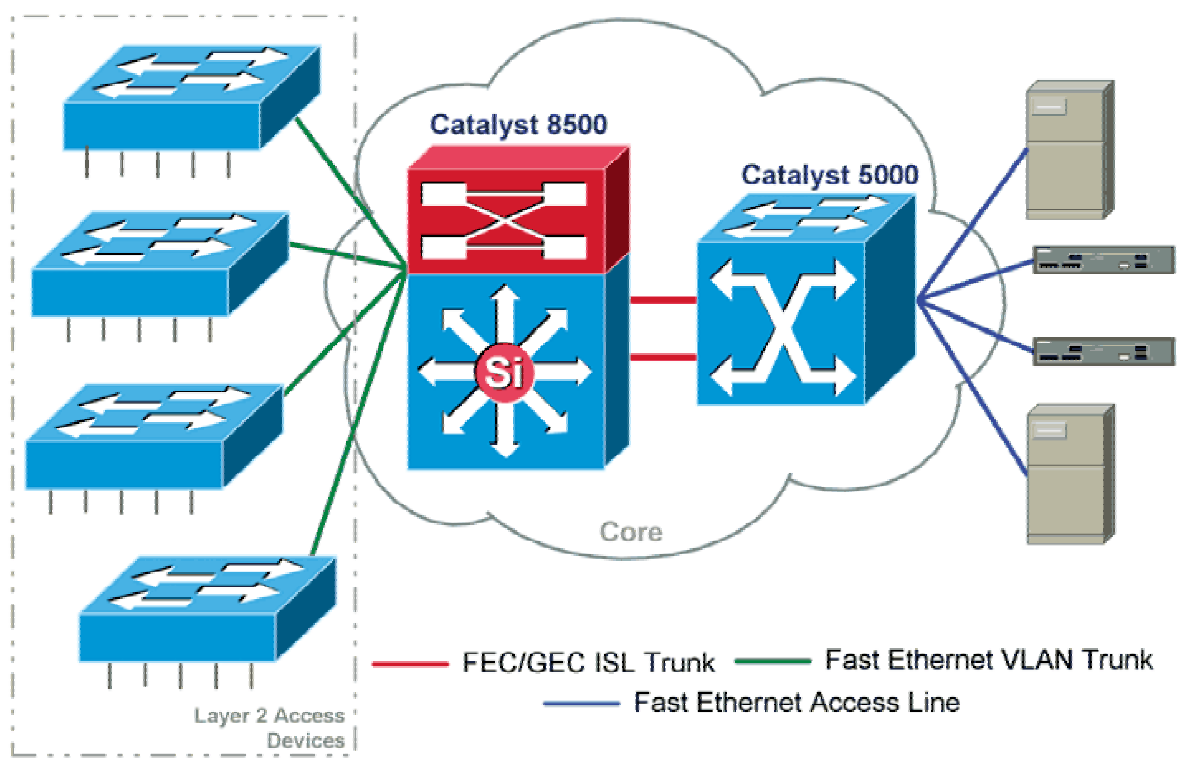

***Fig. 3.*** *Current topology of the central core of the FLNP LAN.*

The fulfillment of the specified works made it possible:

— to increase real throughput of the network by 50-60 % without changing physical interfaces;
— to provide mechanisms of control, analysis and filtration of the network traffic;
— to extend address space (at present, up to 4000 IP-addresses);
— to organize virtual subnetworks for groups of users (or spectrometers) irrespective of their geographical location (in 2002, four subnetworks were created);
— to provide a guaranteed passband for the most important network applications (for example, for concrete spectrometers).

The main objective of further LCN development is to design a fully redundant high-speed core by installing second router 85XX. The availability ot two routing switches and duplicated trunk communication channels will permit a gradual changeover to the Gigabit Ethernet technology. In addition, there will be a possibility of automatic load distribution over the communication channels on the basis of traffic classification. In case of failure of the network core components, the data transfer will be resumed in a few seconds. The failure-resistance in this network structure will be provided: on physical level – by the availability of no less than two connection from each node; on logical level – by using protocols, which make it possible to re-direct data flows if the connection fails.